\documentclass[aps, prd, twocolumn, nofootinbib, floatfix, showkeys, preprintnumbers]{revtex4}
\usepackage{subfigure}
\usepackage{graphicx}
\usepackage{dcolumn}
\usepackage{epsfig}
\usepackage{amsmath}
\usepackage{amsfonts}
\usepackage{amssymb}
\usepackage{color}
\usepackage{hyperref}
\usepackage{dutchcal}
\usepackage{color}
\usepackage{physics}
\usepackage{enumitem}
\usepackage{siunitx}
\usepackage[svgnames]{xcolor}

\hypersetup{
    colorlinks=true,
    citecolor=Teal,
    urlcolor=cyan,
    linkcolor=blue,
}

\makeatletter
\renewcommand{\fnum@figure}{Fig. \thefigure}
\makeatother

\makeatletter
\newsavebox\myboxA
\newsavebox\myboxB
\newlength\mylenA

\newcommand*\xoverline[2][0.75]{%
    \sbox{\myboxA}{$\m@th#2$}%
    \setbox\myboxB\null
    \ht\myboxB=\ht\myboxA%
    \dp\myboxB=\dp\myboxA%
    \wd\myboxB=#1\wd\myboxA
    \sbox\myboxB{$\m@th\overline{\copy\myboxB}$}
    \setlength\mylenA{\the\wd\myboxA}
    \addtolength\mylenA{-\the\wd\myboxB}%
    \ifdim\wd\myboxB<\wd\myboxA%
       \rlap{\hskip 0.5\mylenA\usebox\myboxB}{\usebox\myboxA}%
    \else
        \hskip -0.5\mylenA\rlap{\usebox\myboxA}{\hskip 0.5\mylenA\usebox\myboxB}%
    \fi}
\makeatother

\newcommand\Mbar{\xoverline[0.6]{M}}

\newcommand\Mtilde{\stackrel{\sim}{\smash{M}\rule{0pt}{1.2ex}}}

\DeclareSIUnit\year{yr}

\newcommand{\be}{\begin{equation}}
\newcommand{\ee}{\end{equation}}
\newcommand{\bea}{\begin{eqnarray}}
\newcommand{\eea}{\end{eqnarray}}

\newcommand{\Pl}{\mathrm{Pl}}
\newcommand{\Schw}{\mathrm{Schw}}
\newcommand{\Kerr}{\mathrm{Kerr}}
\newcommand{\bh}{\mathrm{\textsc{bh}}}
\newcommand{\dm}{\mathrm{\textsc{dm}}}
\newcommand{\ekbh}{\textsc{ekbh}}
\newcommand{\cut}{\mathrm{cut}}
\newcommand{\rhorad}{\rho_\mathrm{rad}}
\newcommand{\rhomat}{\rho_\mathrm{m}}
\newcommand{\ext}{\mathrm{ext}}
\newcommand{\Mext}{M_{\ext}}
\newcommand{\eq}{\mathrm{eq}}

\begin{document}

\title{Extremal Kerr Black Hole Dark Matter from Hawking Evaporation}

\author{Quinn Taylor}
\affiliation{Department of Physics/CERCA/Institute for the Science of Origins, Case Western Reserve University, Cleveland, OH 44106-7079 -- USA}
\author{Glenn D.~Starkman}
\affiliation{Department of Physics/CERCA/Institute for the Science of Origins, Case Western Reserve University, Cleveland, OH 44106-7079 -- USA}
\author{Michael Hinczewski}
\affiliation{Department of Physics/CERCA/Institute for the Science of Origins, Case Western Reserve University, Cleveland, OH 44106-7079 -- USA}
\author{Deyan P. Mihaylov}
\affiliation{Department of Physics/CERCA/Institute for the Science of Origins, Case Western Reserve University, Cleveland, OH 44106-7079 -- USA}
\author{Joseph Silk}
\affiliation{Institut d'Astrophysique de Paris (UMR7095: CNRS \& UPMC- Sorbonne Universities), F-75014, Paris, France}
\affiliation{William H. Miller III Department of Physics and Astronomy, The Johns Hopkins University, Baltimore MD 21218, USA}
\affiliation{ Beecroft ~Institute~ of~ Particle~ Astrophysics~ and Cosmology, Department~ of~ Physics,
 University~ of~ Oxford, Oxford OX1 3RH, UK}

\author{Jose de Freitas Pacheco}
\affiliation{Universit\'e de la C\^ote d'Azur - Observatoire de la C\^ote d'Azur, 06304 Nice Cedex - France}

\date{\today}

\begin{abstract}\noindent
The Hawking process results in a monotonic decrease of the black hole mass, but a biased random walk of the black hole angular momentum. 
We demonstrate that this stochastic process leads to  a  significant fraction of primordial black holes becoming extremal Kerr black holes (\textsc{ekbh}s) of one to a  few Planck masses
 regardless of their initial mass.  For these \textsc{ekbh}s, the probability of ever absorbing a photon or other particle from the cosmic environment is small, even in the cores of galaxies.
 Assuming that \textsc{ekbh}s are stable, they behave as cold dark matter, and can comprise all of the dark matter if they are formed with the correct initial abundance.
\end{abstract}


\maketitle

\section{Introduction}
\label{sec:intro}
\noindent
Black holes (\textsc{bh}s) have long been considered a promising class of dark matter candidates \cite{carr_black_1974, carr_primordial_1975, carr_observational_2023, carr_primordial_2016, carr_primordial_2020, green_primordial_2021, hektor_constraints_2018, macgibbon_can_1987, pacheco_quasi-extremal_2023, sasaki_primordial_2018, chapline_cosmological_1975}.
To constitute all the dark matter, they must be formed before Big Bang Nucleosynthesis (\textsc{bbn}), i.e. before the temperature of the Universe has fallen 
below \(\sim \SI{1}{\mega\eV}\), and endure from their epoch of formation to the present \cite{chapline_cosmological_1975, green_primordial_2021, carr_primordial_2018}.
They must  be stable, or at least have lifetimes much longer than the current age of the Universe.

Given evidence for the role of dark matter in the Universe back to the epoch of \textsc{bbn}, plausible candidates include \textsc{bh}s formed in the early Universe, with masses therefore unrelated to those of the stellar mass black holes that emerge from core-collapse supernovae.
Most recent attention has focused on \textsc{bh}s of large mass, \(M \gg \SI{e15}{\gram}\) because they are expected to be stable, with decay lifetimes much longer than the current age of the Universe \cite{carr_primordial_2016}.

As first pointed out in \cite{macgibbon_can_1987}, if \textsc{bh} evaporation leaves behind Planck mass stable relics, then they could be the dark matter.
Barrow {\it et al.} discussed \cite{Barrow:1992hq} this possibility of Planck relics from \textsc{bh} evaporation and concluded that a substantial relic density could remain if the initial mass function of primordial \textsc{bh}s was relatively narrow, and this was followed up by others (e.g. \cite{carr_black_1994}).
Others argued more recently \cite{Alexeyev:2002tg, Chen:2002tu, Nozari:2005ah} that \textsc{bh} decay inevitably resulted in a Planck relic.

In this paper we argue that the Hawking process itself is likely to ``strand'' a small fraction of all primordial \textsc{bh}s as extremal Kerr black holes (\ekbh), which are expected to be stable. 
Recently, Dai and Stojkovic \cite{dai_separating_2023} have argued that \ekbh s  undergo super-radiance and therefore are not stable, even though they do not undergo the Hawking process. 
We comment on this possibility below. 
One way or another, the scenario we present should be instructive for how \textsc{bh} decay even in semiclassical General Relativity may result in naturally Planck relic dark matter if Planck relics are stable.

\textsc{bh}s are expected to decrease in mass via the Hawking evaporation process \cite{hawking_particle_1975, page_particle_1976, firouzjaee_particle_2016, arbey_blackhawk_2019, carter_charge_1974, dai_analytic_2010, macgibbon_quark-_1990, page_hawking_2005, taylor_evaporation_1998, arbey_evolution_2020}.
For a Schwarzschild (i.e. uncharged, non-rotating) \textsc{bh} of mass \(M_{\bh}\), the Hawking temperature is 
\begin{equation}
	\label{eqn:TempSchwarzschild}
	T_{\Schw} = \frac{M_{\Pl}^2}{8\pi M} \simeq \SI{1.1e13}{\giga\eV}\, \frac{\si{\gram}}{M}
\end{equation}
(in units where \(\hbar = c = k = 1\)) and the horizon radius is
\begin{equation}
	\label{eqn:RSchwarzschild}
	R_{\Schw} = \frac{2 M}{M_{\Pl}^2} \,,
\end{equation}
resulting in a Hawking luminosity (for one massless degree of freedom)
\begin{equation}
	\label{eqn:LSchwarzschild}
	{\mathcal L}_{\Schw} = 4\pi R_{\Schw}^2 \, \sigma \, T_{\Schw}^4 = \frac{M_{\Pl}^4}{15360 \pi M^2} \,,
\end{equation}
and a Hawking lifetime of 
\begin{equation}
	\label{eqn:lifetimeSchwarzschild}
    t_{\Schw} = \frac{5120 \, \pi^3 M^3} {M_{\Pl}^4} = 4 \left(\frac{M}{10^{9}\si{g}}\right)^{\!3} \si{\s}\,.
\end{equation}

For Schwarzschild black holes formed in the early Universe to still be around  after \(\SI{13.8}{\giga\year}\), we require \(M_{\Schw} \gtrsim \SI{e15}{\gram}\), placing an apparent lower limit on the mass of the black holes that can serve as dark matter. 
In fact, the limit is somewhat stronger than this, 
\begin{equation}
	\label{eqn:stableMSchwarzschild}
	M_{\Schw} \gtrsim  \SI{e17}{\gram},
\end{equation}
since otherwise the diffuse extragalactic gamma-ray background caused by the Hawking evaporation of such black holes would have been detected \cite{arbey_constraining_2020, hektor_constraints_2018, musco_threshold_2019, katz_femtolensing_2018, niikura_microlensing_2019, carr_cosmic_2021, fernandez_unraveling_2019, nakama_limits_2018, tada_primordial_2015, trimble_existence_1987}.\footnote{
    One might have wondered whether this limit would have been rendered moot by the accretion of ambient particles in the early Universe automatically driving the black hole mass above this limit, however, the accretion process is remarkably ineffective for small black holes, given that \(R_{\Schw}\propto M\).
}

The temperature of a black hole is not, however, determined exclusively by its mass \cite{adler_generalized_2001, wald_general_1984, belgiorno_black_2004, bekenstein_black_1973}.
For a rotating (but uncharged) Kerr black hole with angular momentum \(J\),  the temperature is
\begin{equation}
	\label{eqn:TempKerr}
	T_{\Kerr} = \frac{M_{\Pl}^2}{4\pi M} \frac{\sqrt{1-a_*^2}}{1+\sqrt{1-a_*^2}}\,,
\end{equation}
where 
\begin{equation}
	a_* \equiv \frac{J M_{\Pl}^2}{M^2} \,.
\end{equation}
If the Hawking process involved only photons (they predominate), was entirely in the orbital angular momentum zero channel (it predominates), and was unbiased as to whether each photon was emitted spin-aligned or anti-aligned with the black-hole spin (it is not, but we will return to this point), then the evolution of \(J\) would be an unbiased random walk with step-size 1 (i.e., \(\hbar\)). 

We can investigate the fate of a black hole under this simplified assumption -- does it remain Schwarzschild like if it begins uncharged and with \(J = 0\).
Since \(T_{\Schw}\ll M\) for \(M\gg M_{\Pl} \simeq \SI{2.17e-5}{\gram}\), a radiating Schwarzschild black hole will emit \(N\) particles
\begin{equation}
   N(M) \simeq \frac{1}{\gamma} \int_{M_i}^M \frac{-\dd M}{T_{\Schw}} 
   = \frac{4\pi}{\gamma M_{\Pl}^2} \left(M_i^2 - M^2\right)\, ,
\end{equation}
where \(M_{i}\) is the black hole's initial mass, \(M\) is the black hole's current mass, and
\begin{equation}
    \gamma  = \frac{\pi^4}{30 \zeta(3)} \approx 2.70\,.
\end{equation}
In the meantime, the black hole's spin will undergo an unbiased random walk, and the 
expected root mean square angular momentum will grow to 
\begin{equation}
    \label{eqn:Jsqexpected}
   \left\langle J^2 \right\rangle^{1/2} = \sqrt{N(\epsilon)} =
   \frac{2}{M_\Pl}\sqrt{\frac{\pi}{\gamma}}\left(M_i^2 - M^2\right)^{1/2}
\end{equation}

A black hole becomes an \ekbh \ once \(a_\star = 1\) or when \(J M_{\Pl}^2 = M^2\). Using Eq.~\eqref{eqn:Jsqexpected}, and for \(M_i \gg M_{\Pl}\), this occurs when 
\(M \simeq 1.5 \, \sqrt{M_i M_{\Pl}}\).
In other words, the typical Schwarzschild black hole would be expected to become an \ekbh~ by the time its mass falls from some initial value \(M_{i}\) to \(M \simeq \sqrt{M_{i}  M_{\Pl}}\)!

In order for these \ekbh s to be stable against Hawking radiation before \textsc{bbn}, we would require \(t_{\Schw} \lesssim \SI{1}{\s}\), implying \(M_i \lesssim \SI{e7}{\gram}\).
Essentially every single Schwarzschild black hole ever formed with initial mass between \(M_{\Pl}\) and \(\SI{e7}{\gram}\) would have become an \ekbh~ by \(\SI{1}{\s}\) under this assumption of unbiased emission of photons.

Hawking particle emission is {\em not} however expected to be unbiased with respect to the alignment between the spin of the photon and the spin of the black hole.
Rather, the emission {\em is} calculated to be biased to prefer the emission of photons with spins parallel to that of the black hole \cite{hawking_particle_1975, dai_analytic_2010}. This will reduce the probability of a black hole evolving to extremality.

According to \cite{nomura_black_2013}, for \(a_{*} \ll 1\), the probabilities of emitting photons (in the {\it s}-state) with spin (and thus angular momentum) aligned/anti-aligned with \(J\) are
\begin{equation}
    \label{eqn:Psmallastar}
    P_{{\uparrow \downarrow}}(a_*) = \frac{1}{2} \mp a_{*} + {\cal O}(a_{*}^2). 
\end{equation}
To evaluate the efficacy of this bias in suppressing the formation of \ekbh, we must extend these approximate formulae to \(\vert a_{*} \vert = 1\).  
Consider
\begin{equation}
	\label{eqn:Pofastar}
     P_{{\uparrow \! \uparrow \atop \uparrow \! \downarrow}}(a_*) = \frac{1}{2} \mp a_* \pm \frac{a_*\vert a_*\vert}{2}	\,.
\end{equation}
This correctly reduces to Eq.~\eqref{eqn:Psmallastar} for \(\vert a_{*} \vert \ll 1\), is monotonic on \(a_{*} \in [-1,1]\), and gives \(P_{{\uparrow \! \uparrow \atop \uparrow \! \downarrow}}(\mp a_{*}) = 0\) as required.

Of course, a Kerr black hole doesn't just emit photons but will emit all particles with masses \(m < T_{\Kerr}\). 
They will similarly preferentially be emitted in the {\it s}-wave, and with spins preferentially aligned with the angular momentum of the black hole \cite{hawking_particle_1975, dai_analytic_2010}.
For simplicity we focus on the {\it s}-wave photons with the expectation that we will correctly capture the qualitative behaviour.

The authors of \cite{arbey_evolution_2020, sorce_gedanken_2017, belgiorno_black_2004} have studied the effect of this spin-down bias on the expected evolution of \(a_{*}\).
They conclude that even black holes with initial \(\vert a_{*} \vert\) very near \(1\) would be expected to evolve to very near \(a_{*} = 0\) once Hawking evaporation becomes significant.   
However, the question we address here is not the expected evolution of the black hole, but the  evolution of a large number of black holes governed by the stochastic nature of the Hawking process.
Unsurprisingly,  spin-down bias substantially reduces the probability of reaching extremality, but by how much? 
Do enough black holes become \ekbh s for these to constitute the dark matter, and what would we expect their mass distribution to be?

\section{The Distribution of \ekbh}
\noindent
To answer these questions, we simulated the evolution of a large number of initially Schwarzchild black holes as they evaporated through the Hawking process. 
Starting from an initial mass of \(M(t=0)\gg M_{\Pl}\), and \(J(t=0) = 0\), each black hole was followed until either:
\begin{enumerate}[label=\alph*.]
    \item the mass of the \textsc{bh} reaches some cut-off value \(M_{\cut}\)~\cite{chen_black_2003, chen_inflation_2005, adler_generalized_2001};
    \item the black hole becomes extremal i.e. \(|J| \geq (M/M_{\Pl})^2\) (at which point we set \(|J|=(M/M_{\Pl})^2\)).
\end{enumerate}
We assume that the terminal \(M = M_{\cut}\) black holes will decay away, and focus on the properties of the \ekbh\footnote{
    The code can be found at \url{https://github.com/qtaylorphys/BH_Extremal}.
    The outline of the code is as follows: At each time step \(t \to t+\delta t\), the mass \(M(t)\) is updated via \(M(t+\delta t) = M(t) -\delta M(t)\), where \(\delta M(t) = x(t) \, T_{\Kerr}(M(t), a_*(t))\). Here \(x(t)\) is a random value sampled from the Planck distribution \(p(x) = (1 / 2 \zeta(3)) \, x^2/(e^x -1)\), and \(T_{\Kerr}\) is given by Eq.~\eqref{eqn:TempKerr}.
    Meanwhile, the angular momentum \(J\) of the \textsc{bh} is changed by \(\pm 1\) with probabilities given by Eq.~\eqref{eqn:Pofastar}.
    We have checked that our results don't depend materially on whether \(P_{\uparrow\!\uparrow / \uparrow\!\downarrow} (a_*)\) is calculated using \(M(t)\) or \(M(t + \delta t)\).}.
    
As expected the black holes remained nearly Schwarzchild (i.e. \(a_{*}^{2} \ll 1\)) for the majority of their evolution until \(M\) fell to just a few times \(M_{\Pl}\), i.e. until the the last few  Hawking particle emissions. 
Of the black holes that we simulated, approximately 22\% of them became extremal, independent of the value of \(M_{0}\), so long as \(M_{0} \gg M_{\Pl}\).

\(P(\Mext)\), the probability density of a black hole becoming extremal at mass \(\Mext\), is well-approximated by 
\begin{equation}\label{eqn:finalmdf}
\begin{split}
    &P(\Mext) = K \Mext \sqrt{\nu}
    \left(1 - \frac{\lfloor \Mext^{2} \rfloor}{\Mtilde^{2}}\right)^{\!\!2}
    \!\exp(\frac{-\nu \lfloor \Mext^2\rfloor^2}{M_{\Pl}^{4}})
\end{split}
\end{equation}
where
\begin{subequations}
\begin{align}
    K &= \frac{8\sqrt{\pi}}{\gamma}, \\
    \nu &= \ln \Biggl[ 1 + \frac{2M_{\Pl}^2}{\Mtilde^2} - \frac{M_{\Pl}^4}{\Mtilde^4} \Biggr], \\
    \Mtilde &= \frac{1}{2} \Biggl(\Mext + \sqrt{\Mext^2 + \frac{\gamma}{2\pi} \, M_{\Pl}^2}\,\Biggr) \,
\end{align}
\end{subequations}
and \(\lfloor x \rfloor\) is the greatest integer \(\leq x\).
Black holes with \(\sqrt{n-1} \leq \Mext / M_{\Pl} < \sqrt{n}\) have \(J = n\), for \(n~\!\!\in~\!\!\mathcal{Z}^{>1}\).

\begin{figure}
    \centering
    \includegraphics[]{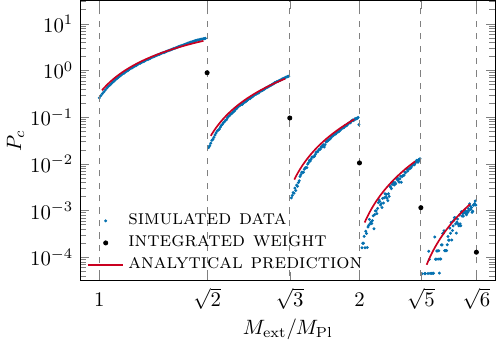}
    \caption{The mass distribution function \(P_c(\Mext)\) (scale on left vertical axis) of extremal Kerr black holes emerging from the Hawking-process evolution of an ensemble of initially Schwarzshild black holes of mass \(M_{0} \gg M_{\Pl}\) (allowing only for {\it s}-wave photon emission).
    Numerical results (blue dots) are well described by Eq.~\eqref{eqn:finalmdf} (red curve), which is the result of an approximate analytic treatment of the evolution of the ensemble (see Appendix~\ref{app:Analytic})).
    Black dashed vertical lines mark \(\Mext / M_{\Pl} = \sqrt{n}\), for \(n \in \mathcal{Z}^{>0}\), and the black holes with \(\sqrt{n-1} \leq \Mext / M_{\Pl} < \sqrt{n}\) have \(J = n\). 
    Since consequently \(a_{*}^{2} > 1\) except at \(\Mext / M_{\Pl} = \sqrt{n}\), which is unphysical, we also plot the integrated weight between \(\sqrt{n-1}\) and \(\sqrt{n}\) and display the values as black dots.}   
    \label{fig:Hist_with_Fits}
\end{figure}

The final distribution of \ekbh~masses \(P_c(\Mext)\), where 
\begin{align}
P_c(\Mext) = \frac{P(\Mext)}{\int_{1}^{\infty} P(\Mext') \, \dd\Mext'}
\end{align}
is shown in Fig. \ref{fig:Hist_with_Fits} for \(M_{\cut} = M_{\Pl}\).
The jagtooth shape of the distribution seems extraordinary, however in Appendix~\ref{app:Analytic} we show that this can be understood as emerging from the doubly stochastic process of black hole mass loss and black hole angular momentum-biased random walk.

We see that the distribution is heavily weighted toward the lightest possible \ekbh, and so the fraction of initially Schwarzschild black holes that reach Kerr extremality is a steeply decreasing function of \(M_{\cut}\).    
The average mass of the \ekbh~is \(\Mbar \simeq 1.3 \, M_{\Pl}\).

\begin{figure}[h]
    \centering
    \includegraphics{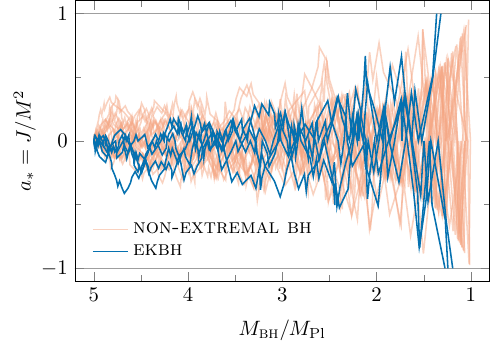}
    \caption{The evolution of \(a_{\star}\) for 25 black holes, each with an initial mass of \(5 M_{\Pl}\), as they evaporate and either become a \ekbh (blue) or reach \(M = M_{\Pl}\) (orange).}
    \label{fig:cross}
\end{figure}

Of note is the structure of the distribution of final masses. 
The distibution is a sawtooth that monotically increases toward a local maximum as \(M\) increases from \(\sqrt{n-1} \, M_{\Pl}\) to \(\sqrt{n} \, M_{\Pl}\) (for integers \(n \geq 2\)), then falls vertically to a lower value for the next sawtooth maximum. 
This distribution can be explained by considering a black hole in the last time step before it becomes extremal. 
With each Hawking photon emitted, the value of \(J\) changes by \(\pm 1\) while the mass changes by an amount drawn randomly from the Planck distribution of the appropriate temperature. 
With the final photon, the black hole spin changes from \(J = n-1\) to \(J = n\) (in the spin up case), and the black hole becomes extremal with a mass \(M \leq \sqrt{n} \, M_{\Pl}\). 
The evolution of \(a_{\star}\) as a function of \(M\) is shown in Fig.~\ref{fig:cross}.

Because we have quantized \(J\) but not \(M_{\bh}\), the black holes with \(\sqrt{n-1} \leq M_{\bh} / M_{\Pl} < \sqrt{n}\) and \(J = n\) have \(a_{*}^{2} > 1\). 
This is unphysical, and reflects the need to properly deal with the quantum mechanical nature of the decay. 
One way to resolve this issue
\cite{deFreitasPacheco:2020wdg} is to take the final mass of the black hole to be \(\sqrt{n} \, M_{\Pl}\) so that \(a_{*} = \pm 1\) exactly. 
For illustrative purposes we therefore also plot in Fig.~\ref{fig:Hist_with_Fits} the integrated weight under each ``tooth'' of the mass distribution function, which corresponds to this resolution (ignoring the low-likelihood possibility that the final photon jumps the black hole across an entire tooth).

\begin{figure}[h]
    \centering
    \includegraphics[]{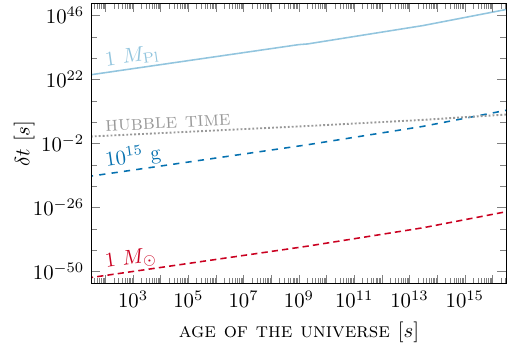}
    \caption{The time \(\delta t\) between \textsc{cmb} photons hitting different mass \ekbh s as a function of the age of the Universe (solid curves), compared to the Hubble time (dotted curve).}
    \label{fig:hitting_times}
\end{figure}

While the final black holes are extremal and no longer radiating, if they were to absorb any energy they would be kicked out of extremality, begin to radiate again, and would once again be more likely to decay away than return to extremality. 
This would be a problem if these black holes were of a much larger mass, but with a geometric cross section \(\sigma = 16 \pi M_{\bh}^2 / M_{\Pl}^4\) these black holes are extremely weakly interacting, and are unlikely to ever be knocked out of extremality once they reach it \cite{taylor_how_2022}. 
For example, the last time that a typical primordial \(\SI{e9}{\gram}\) Schwarzschild black hole would have even encountered a cosmic microwave background (\textsc{cmb}) photon was well before recombination \cite{taylor_how_2022}; see Fig.~\ref{fig:hitting_times}.

\section{\ekbh s as Dark Matter}
\noindent
The formation of black holes during the early Universe has been a rich area of study with many proposed mechanisms \cite{carr_primordial_2021, deng_primordial_2017, fonseca_primordial_2012, garcia-bellido_primordial_2017, gow_primordial_2023, hawking_black_1989, hawking_bubble_1982, inomata_double_2018, kannike_single_2017, nakama_identifying_2014, polnarev_formation_1991, yokoyama_formation_1995, jedamzik_could_1998, kawasaki_primordial_2013, garcia-bellido_gravitational_2016}. 
One common mechanism for the formation of these primordial black holes (\textsc{pbh}s) comes from density perturbations in the early Universe \cite{carr_primordial_1975, dvali_new_2004, evans_observation_1994, garcia-bellido_density_1996, hawking_gravitationally_1971, kuhnel_effects_2016, kuhnel_ellipsoidal_2016, musco_threshold_2019, musco_computations_2005, musco_primordial_2009, musco_threshold_2021, nakama_identifying_2014, niemeyer_near-critical_1998, polnarev_curvature_2007, shibata_black_1999}. 
A black hole will form if the density of the perturbed region reaches some critical density \(\delta_{c}\). The resulting \textsc{bh} mass will be proportional to the horizon mass \cite{green_primordial_2021}
\begin{equation}
    \label{eqn:mhorizon}
    M_H\sim 5 \times 10^{20} M_{\Pl}\left(\frac{t}{\SI{e-23}{\s}}\right)
\end{equation}

In order for \ekbh s to be the dark matter, and in order not to disturb \textsc{bbn}, they need to have been in place before the temperature of the Universe has fallen to \(\SI{1}{\mega\eV}\), when \(M_H \sim 5 \times 10^{43} M_{\Pl}\). 
However, since a \textsc{bh} will only evaporate through Hawking emission when its Hawking temperature is greater than the ambient temperature \cite{firouzjaee_particle_2016, hawking_soft_2016}, the initial black holes must be light enough that \(T_{H} (M_{i}) \gg \SI{1}{\mega\eV}\), i.e. \(M_i \ll 5 \times 10^{20} \, M_{\Pl}\). 
The upper limit on \(M_{i}\) is lowered even more due to the decay lifetimes \(\tau\) of \ekbh s being constrained to less than 1 second. 
Thus \(M_{i} \leq 10^{13} M_{\Pl}\). 

Naively one might then simply solve for the initial number density of black holes \(n_{\bh}^{i}\) required to constitute all of the dark matter using 
\begin{equation}
    \label{eqn:nBHi}
    n_{\bh}^i = \frac{\Omega_{\dm} \, \rho_c }{P_{\ext} \Mbar} \, (1+z_i)^3 
\end{equation}
where the current average density of dark matter is \(\rho_{\dm} = \Omega_{\dm} \rho_c \simeq \SI{2.5e-30}{\gram\per\cm^{3}}\), \(z_{i}\) is the redshift of \textsc{bh} formation, \(P_{\ext} \approx 0.11\) is the probability of a primordial black hole becoming extremal, and \(\Mbar \approx \SI{e-5}{\gram}\) is the average mass of the \ekbh~that ultimately remains after \textsc{bh} decay. 
For \ekbh s made at \(t = \SI{e-22}{\s}\) after the Big Bang, the initial number density of black holes is \(n_{\bh}^{i} = \SI{2.27e12}{\cm^{-3}}\). 
Before these black holes become extremal, however, they radiate away almost all their mass and contribute to the overall radiation energy density of the Universe \footnote{For simplicity, we will take the number of effective relativistic degrees of freedom of the Universe to always be equal to the number for the Standard Model of particle physics at high temperature, \(106.75\).}.
\begin{equation}
    \dv{\rhorad}{t} = -\dv{M_{\bh}}{t} \, n_{\bh}^{i} \!\left(\frac{a}{a_i}\right)^{\!\!-3} - 4 \rhorad \!\left(\frac{\dot{a}}{a}\right)
    \label{eqn:drhodt}
\end{equation}
where 
\begin{equation}
    -\dv{M_{\bh}}{t} = \sigma A_{\bh} T_{\bh}^4
\end{equation}
is the energy radiated away from the black hole. 

Because decaying \textsc{bh}s remain nearly-Schwarzschild (i.e. \(a_{*}^{2} \ll 1\)) throughout their lifetime until nearly their final moments, and because each Hawking particle typically reduces the mass by only a small fraction (i.e. \(T_{\bh} \ll M_{\bh}\)), we can approximate the time evolution of the \textsc{bh} mass as nearly the expected evolution of the mass.
For a Schwarzschild \textsc{bh}, this gives
\begin{equation}
    \label{eqn:MBHoft}
    M_{\Schw}(t) = \left[ M_{\Schw}^3(t_i) - \frac{M_{\Pl}^4}{5120\pi}(t-t_i) \right]^{1/3}
\end{equation}
To find \(\rho(t)\), we utilize the Friedmann equation
\begin{equation}
    \left(\frac{-\dot{T}}{T}\right)^2 = \frac{8\pi}{3 M_{\Pl}^2}\left(\rhorad + \rhomat\right),
    \label{eqn:Friedmann_eq}
\end{equation}
where we used \(a / a_{0} = T_{0} / T\). Taking the time derivative of both sides 
\begin{equation}
    \label{eqn:Tdiffeq}
    \frac{\ddot{T}}{T} - 3\left(\frac{\dot{T}}{T}\right)^2 = -\frac{4\pi}{3 M_{\Pl}^2} \, M_{\bh}(t) \, n_{\bh}^{i} \!\left(\frac{T}{T_i}\right)^3 \,.
\end{equation}

Solving Eq.~\eqref{eqn:Tdiffeq} numerically gives \(T(t)\) for \(t < \tau\), enabling us to find \(\rhorad(t=\tau)\), as a function of the initial mass and number density of the black holes.
(For simplicity we assume that all the black holes were formed at the same time with a unique mass.)
\begin{equation}
    1 = \frac{\rhomat(t_{\eq})}{\rhorad(t_{\eq})} \simeq \frac{\rho_{\bh}(t=\tau)}{\rhorad(t=\tau)} \frac{1+z_\tau}{1+z_{\eq}}
    \label{eq:rho_equality}
\end{equation}
where we have substituted in \(\rho_{\bh}\) for \(\rhomat\) since the matter energy density is dominated by the energy density from \textsc{dm}. Eq. \eqref{eq:rho_equality} is then used to find the values for \(n_{i}\) that satisfy the equality. 

\begin{figure}[h]
    \centering
    \includegraphics{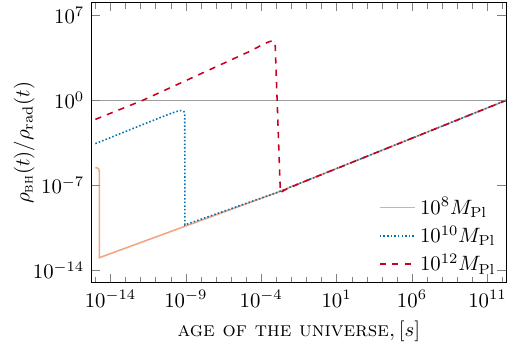}
    \caption{The evolution of \(\rho_{\bh}(t) / \rhorad (t)\) as black holes formed at \(t_{i} = \SI{e-15}{\s}\) with an initial mass \(M_{i}\) evaporate. Here \(\rhorad\) is taken to be the radiation energy density in a radiation-dominated Universe (with the 106.75 effective relativistic degrees of freedom of the Standard Model at high temperature).}
    \label{fig:rho_rho}
\end{figure}

For a single set of black holes made at time \(t = \SI{e-15}{\s}\), the behavior of \(\rho_{\bh} / \rhorad\) as the black holes evaporate is shown in Fig.~\ref{fig:rho_rho}.

\section{Conclusions and Considerations}
\noindent
We have shown that the biased random walk of a black hole's angular momentum can result in extremal Kerr black holes when the mass of these black holes is initially of order \(M_{\Pl}\). 
If the minimum mass of a stable black hole is \(M_{i} < 10^{13} \, M_{\Pl}\), then of the initial number of black holes simulated around 22\% become extremal Kerr black holes with average mass \(\bar{M} = 1.3 \, M_{\Pl}\). Using this percentage and average mass, we have calculated the initial density, as a function of initial mass, of \ekbh s needed to be the dark matter in the Universe today. 

For our analysis, however, we have  considered that the contribution of the black hole entropy is only due to the outer horizon of the Kerr black holes. 
References \cite{volovik_effect_2021, singha_hawking_2023} show that the inner horizon of black holes effects the entropy of the black hole, ultimately resulting in the temperature of a black hole being only proportional to mass. 
This will make all black holes with inner horizons radiate like a Schwarzchild black hole, making them unstable. 
In \cite{dai_separating_2023}, it is argued that waves incident on black hole horizons undergo super-radiance and take away energy from the black hole. 
This would also cause the \ekbh~ to radiate and make them unsatisfactory as dark matter candidates.  
However, the calculation of \cite{dai_separating_2023} assumed a single isolated \ekbh, whereas dark matter of mass \(\mu \, M_{\Pl}\) would have a cosmological average abundance of \(2/\mu \times \SI{e-19}{\cm^{-3}}\), and thus a mean separation of just \(\sim \SI{15}{\km}\). 
Their individual Kerr metrics would dominate the instantaneous local geometry at most over that distance.  
The prospect that the dark matter is stabilized by its own many-body effects is intriguing.  

While \ekbh s are a good candidate for the dark matter in the Universe, they are a troubling candidate. 
Their small geometric cross-section is a double-edged sword: while this ensures the stability of \ekbh's due to their low interaction rate with other objects in the Universe, it also means that detecting such dark matter would also be extremely difficult, though perhaps not impossible\cite{lehmann_direct_2019, bai_primordial_2020,alexandre_new_2024, pacheco_primordial_2020, pacheco_quasi-extremal_2023}.

\acknowledgements
G. D. S. and Q. T. acknowledge support from DOE grant DESC0009946; Q.T. from the GEM Fellowship. D.~P.~M. acknowledges support from NASA ATP grant RES240737. This work made use of the High Performance Computing Resource in the Core Facility for Advanced Research Computing at Case Western Reserve University.

\bibliography{main}
\section*{Appendices}
\appendix
\renewcommand{\thesection}{\Alph{section}}

\section{Approximating the Extremal Mass Distribution\\[-1em]}
\label{app:Analytic}
\noindent
We want to approximate analytically the distribution of extremal black hole masses that emerges from the numerical evolution of an ensemble of \ekbh's (blue dots in Fig.~\ref{fig:Hist_with_Fits}). 
The evolution of a black hole is driven by two stochastic processes: the random decrease in the black hole's mass, and the biased random walk of the black hole's angular momentum.\footnote{
   In principle, we should add a third random process -- the random walk of the black hole's electromagnetic charge. 
   Indeed, to really understand the formation of extremal black holes near the Planck mass we might rightly consider the random walk of all the black hole's \(SU(3) \times SU(2) \times U(1)\) gauge charges, however we defer those complications to future considerations.
}

Note that for convenience, in this appendix we set \(M_{\Pl} = 1\) in addition to \(\hbar = c = k = 1\), and take the minimum mass of a black hole to be \(M_{\cut} = 1\).

\subsection*{Approximating the mass trajectory}
\noindent
We start by defining an affine parameter \(\lambda\) that increases monotonically as the black hole decays.  
The black hole starts out at \(\lambda = 0\) with mass \(M(0) = M_{0}\) and angular momentum \(J(0) = 0\).

Our first approximation is related to the evolution of the mass of the black hole \(M(\lambda)\). 
Though this evolution is stochastic in the simulations, \(M (\lambda)\) stays in the vicinity of a deterministic curve \(\Mbar (\lambda)\), with only small discrepancies until near the end of the evolution. 
For the majority of the black hole's Hawking-process evolution, \(M (\lambda) \gg 1\), hence \(T (\lambda) \ll M (\lambda)\). 
Also  \(|a_{\star} (\lambda)| = |J(\lambda) / M^{2} (\lambda)| \ll 1\), so the black hole is nearly Schwarzschild.

The individual decrements in \(M (\lambda)\), \(\delta M (\lambda)\), are each stochastic draws from a black body distribution of temperature \(T (\lambda)\).
However, since a large number of these decrements are needed to substantially decrease \(M (\lambda)\), the stochasticity of \(\delta M (\lambda)\) gets averaged out. 
We can thus replace the stochastic mass decrement by a deterministic Schwarzschild approximation:
\begin{equation}
    \label{eqn:Mbaroft}
    \delta \Mbar(\lambda) \approx \gamma \xoverline{T}(\lambda) \,,
\end{equation}
where
\begin{align}
    \gamma = \frac{\pi^4}{30 \zeta(3)} \qquad \text{and} \qquad \xoverline{T}(t) = \frac{1}{8 \pi \Mbar(t)}.\nonumber
\end{align}

This deterministic approximation Eq.~\eqref{eqn:Mbaroft} for the black hole mass decrement allows us to define \(\lambda\) in such a way that \(\Mbar (\lambda)\) obeys a simple differential equation:
\begin{equation}
    \label{eqn:Mbaroftdiffeq}
    \frac{\dd\Mbar(\lambda)}{\dd \lambda} = \frac{\gamma}{8 \pi \Mbar (\lambda)},
\end{equation}
which has the solution
\begin{equation}\label{eqn:barm}
\Mbar(\lambda) = M_0 \sqrt{ 1 -\frac{\gamma\lambda}{4\pi M_0^2} }\,.
\end{equation}
We will always restrict ourselves to \(\lambda < 4 \pi M_{0}^{2} / \gamma\), so that \(\Mbar (\lambda)\) as given by Eq.~\eqref{eqn:Mbaroftdiffeq} is real and positive.

We can easily relate the affine parameter \(\lambda\) to the time \(t\), because for a Schwarzschild black hole Eq.~\eqref{eqn:LSchwarzschild} gives
\begin{equation}
    \dv{\Mbar}{t} = -\frac{1}{15360 \pi}\frac{1}{\Mbar^2}\,.
\end{equation}
Comparing to Eq.~\eqref{eqn:Mbaroft}
\begin{equation}
    \dv{\lambda}{t} = 1920\pi^2 \gamma M_0\sqrt{1 - \frac{\gamma\lambda}{4\pi M_0^2}}\,,
\end{equation}
so that
\begin{equation}
    t(\lambda) = \frac{M_0}{240\pi\gamma^2}\left[1 - \sqrt{1-\frac{\gamma\lambda}{4\pi M_0^2}} \, \vphantom{\frac{\dfrac{a}{a}}{\dfrac{a}{a}}} \right]
\end{equation}

Near the end of the black hole's evolutionary trajectory, the mass decrements are no longer small compared to the black hole mass, and it is no longer necessarily the case that \(|a_{\star} (\lambda)| \ll 1\), indeed for an \ekbh~we have \(|a_\star(\lambda)| \to 1\). 
There are therefore noticeable differences between the actual \(M (\lambda)\) and \(\Mbar (\lambda)\). 
However, in order to enable an analytical approach, we will ignore these differences and assume \(M (\lambda) \approx \Mbar (\lambda)\).

\begin{figure}
    \centering
    \includegraphics{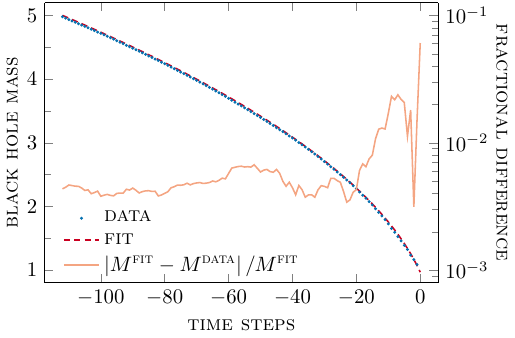}
    \caption{The average mass values of 488 \ekbh~trajectories plotted against Eq.~\eqref{eqn:barm}.}
    \label{fig:trajectories}
\end{figure}

\subsection*{Trajectory end time distribution}
\noindent
Unlike the evolution of the black holes mass, the evolution of the angular momentum cannot be approximated as deterministic. 
With each photon emission, the spin \(J\) changes by \(\pm 1\) with probability given by \eqref{eqn:Pofastar}, which we can rewrite in terms of the probability to emit a photon with spin up (\(+1\)) or down (\(-1\)):
\begin{align}
    \rho_{\uparrow}(J,M)  &= \begin{cases} \dfrac{(J-M^2)^2}{2M^4} & J\ge 0\\[10pt] \dfrac{(J-M^2)^2-2J^2}{2M^4} & J < 0 \end{cases}
    \label{eqn:rhoupsimplified}\\
    \rho_{\downarrow}(J,M) &= 1 - \rho_{\uparrow}(J,M)
\end{align}
In the simulations, the black hole evolutionary trajectory starts with \(J = 0\) at \(\lambda = 0\) and ends at some \(\lambda_f\) when \(|J| \geq M^{2}\) (or  \(M = 1\)).
We can study this in the approximation that \(M\) changes deterministically as \(\Mbar(\lambda)\).
We can model the terminal-\(\lambda\) distribution with a survival probability, \(\Sigma (\lambda)\), or the cumulative probability that the spin trajectory has not reached \(\lambda_{f}\).
This is related to the probability density \({\cal P} (\lambda)\) of trajectory termination \(\lambda\), i.e. \(\lambda_{f}\), by
\begin{equation}
    \Sigma (\lambda) = 1 - \int_0^{\lambda} \dd{\lambda'} {\cal P}(\lambda')
\end{equation}
Note that \(\Sigma (0) = 1\). 
If we find \(\Sigma (\lambda)\), then \({\cal P} (\lambda) = -\dd \Sigma (\lambda) / \dd \lambda\).
We can then change variables to get the probability density at end times of final masses, \({\cal P} (M)\).

The survival probability satisfies
\begin{equation}
    \dv{\Sigma (\lambda)}{\lambda} = -k(\lambda) \, \Sigma(\lambda),
\end{equation}
where \(k (\lambda)\) is the rate of loss, or the rate at which a trajectory reaches the end conditions at \(\lambda\). In other words, the fraction of surviving trajectories will be diminished by \(k (\lambda) \Sigma (\lambda)\) in times between \(\lambda\) and \(\lambda + \dd{\lambda}\).
With the boundary conditions specified above,
\begin{equation}\label{eqn:sigsol}
    \Sigma(\lambda) = \exp\left(-\int_0^\lambda \dd\lambda^\prime \; k(\lambda^\prime)\right),
\end{equation}
and hence
\begin{equation}\label{eqn:psol}
    {\cal P}(\lambda) = -\frac{\dd\Sigma(\lambda)}{\dd\lambda} = k(\lambda) \exp\left(-\int_0^\lambda \dd\lambda^\prime k(\lambda^\prime)\right).
\end{equation}

We are only interested in \(\lambda < \lambda_{max} = \left(4 \pi / \gamma \right)(M_0^2 - 1)\), so that \(\Mbar(\lambda) > 1\).
Because only about 21\% of trajectories end up as extremal, we can approximate the exponential term in \eqref{eqn:psol} with \(1\). 

In the code, the condition of extremality is \(|J| \geq M^{2}\), if the final spin jump occurs between increment \(\lambda\) and \(\lambda + \delta \lambda\) the mass that determines \(\rho_{\uparrow}\) is the mass \(M (\lambda) \approx \Mbar(\lambda)\) and not \(\Mbar(\lambda + \delta\lambda)\).
To distinguish these two slightly different masses, we will denote the ultimate mass as \(\Mext\), and the penultimate mass as \(\Mtilde (\Mext)\). 
The expected difference between them is
\begin{align}
    \dd\Mbar(\Mtilde) &= \gamma\overline{T}(\Mtilde) = \frac{\gamma}{8\pi \Mtilde}
\end{align}
Thus
\begin{equation}
    \Mtilde (\Mext) = \Mext + \frac{\gamma}{8\pi \Mtilde}\,.
\end{equation}
so that
\begin{align}
    \Mtilde(\Mext) &= 
        \frac{1}{2}\left(
            \Mext + 
            \sqrt{\Mext^2 
                + \frac{\gamma}{2\pi}}
         \right)
     \label{eqn:Mtilde}
\end{align}

We can determine \(\lambda(\Mext)\) from \eqref{eqn:barm}, which we write as 
\begin{align}
    \lambda(\Mext) &= \frac{4\pi}{\gamma}\left(M_0^2-\Mext^2\right)-1.  \label{eqn:tm}
\end{align}

Using the change of variable theorem for probability densities, the distribution of final masses \(P (\Mext)\) is:
\begin{equation}\label{eqn:pmf}
    P(\Mext) = \left|\frac{\dd\lambda}{\dd\Mext}\right| {\cal P}(\lambda(\Mext)) \approx \frac{8 \pi \Mext}{\gamma} \, k(\lambda(\Mext)).
\end{equation}

\begin{figure*}
    \centering
    \includegraphics{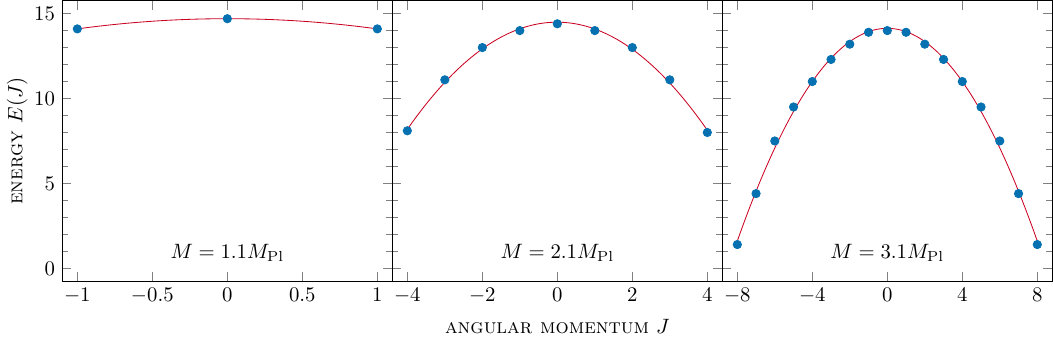}
    \caption{Results of spin dynamics simulations at three different fixed mass values, with reflecting boundary conditions at the smallest and largest possible values of the spin, averaged over both even and odd time steps. The points show \(E(J) = -\log p(J)\), where \(p(J)\) is the distribution of spin values. The curves show parabolic fits to the simulation data.}
    \label{fig:parab}
\end{figure*}

\subsection*{Approximating the trajectory loss rate}
\noindent
There are two major contributions to the loss function \(k (\lambda)\) at late  \(\lambda\): i) the spin \(J = \lfloor \Mext^{2} \rfloor\), and the angular momentum increases \(J \rightarrow J+1\); ii) the spin \(J = -\lfloor \Mext^{2} \rfloor\), and the angular momentum decreases \(J \rightarrow J-1\). 
By symmetry, both of these contributions have the same probability, so we can focus on the positive spin case and multiply by a factor of 2. 

Since \(\Mext\) is slightly smaller than \(\Mtilde\), we know that \(\lfloor \Mext^{2} \rfloor \leq \lfloor \Mtilde^{2} \rfloor\), with the two floor values usually equal. 
However there are narrow ranges of \(\lambda\) when \(\lfloor \Mtilde^{2} \rfloor = \lfloor M^{2}_{f} \rfloor + 1\), and in these ranges \(J\) can go as high as \(\lfloor \Mtilde^{2} \rfloor\). 
During these times, we also have possible transitions from \(J = \lfloor \Mtilde^{2} \rfloor \to \lfloor \Mtilde^{2} \rfloor + 1\) that would end the trajectory (and their negative spin counterparts). 
However, as we will see below, the probability of being at \(J = \lfloor \Mtilde^{2} \rfloor = \lfloor \Mext^{2} \rfloor + 1\) will be much smaller than \(J = \lfloor \Mext^2 \rfloor\), so we can safely ignore the contribution of these transitions.

Thus the loss rate \(k (\lambda)\) can be expressed as:
\begin{equation}\label{eqn:kt}
\begin{split}
    k(\lambda) &= 2 p_\lambda(J=\lfloor \Mext^2 \rfloor) \rho_{\uparrow}(J=\lfloor \Mext^2 \rfloor,\Mtilde)\\
    &= p_\lambda(J=\lfloor \Mext^2 \rfloor) \frac{(\lfloor \Mext^2 \rfloor - \Mtilde^2)^2}{\Mtilde^{4}}.
    \end{split}
\end{equation}
\(\rho_{up}\) is given by \eqref{eqn:Pofastar}, and \(p_\lambda(J)\) is the probability to be at spin \(J\) at affine parameter \(\lambda\). 
The spin dynamics can be seen as a discrete random walk under a force \(\rho_\text{up} - \rho_\text{down}\) that is approximately given by \(-2 J / \Mbar^{2} (\lambda)\) for small \(J\). 
This acts like a Hookean restoring force, biasing the walk toward \(J = 0\).  
Since \(\Mbar^{2}(\lambda)\) decreases with \(\lambda\), the effective spring constant gradually increases.  
Because the change in \(\Mbar (\lambda)\) per  \(\lambda\) increment is small, and the fraction of trajectories lost per \(\lambda\) step is also small, we will assume the system is roughly in quasi-equilibrium. 
In other words, we will approximate \(p_{\lambda} (J)\) by the equilibrium distribution of \(J\) that would have occurred if the mass was fixed at a value \(M = \Mbar (\lambda)\), and the boundary conditions at the smallest and largest allowed spin values were reflecting rather than absorbing.  If the spin dynamics were a continuous Wiener process, this equilibrium distribution under a Hookean force would be a Gaussian.  In the actual system the discreteness of the dynamics complicates matters and prevents a simple closed form solution. However inspired by the continuum case, we will assume a Gaussian ansatz, \(p_{\lambda} (J) \propto \exp(-E_{\lambda} (J))\), where \(E_{\lambda} (J) = E_{\lambda} (0) + \nu (\Mbar (\lambda)) J^{2}\) for some function \(\nu(M)\). 

In Fig.~\ref{fig:parab}, we show simulation results for three different fixed values of the mass, where the points represent \(E(J) = -\log p(J)\) from the numerically determined equilibrium distribution \(p (J)\) (under reflecting boundary conditions).  
In each case a parabola (red curve) provides an excellent fit, validating the Gaussian ansatz.  
Note that the spin dynamics has the property that if \(J = 0\) at \(\lambda = 0\) then only even values of \(J\) are possible at even steps in \(\lambda\), and odd values of \(J\) at odd steps in \(\lambda\).
Thus, strictly speaking, the distribution forever oscillates between even and odd values from \(\lambda\)-step to \(\lambda\)-step. 
We simplify the situation by averaging over both even and odd \(\lambda\)-steps, which gives an approximately Gaussian stationary distribution.

\begin{figure}[h]
    \centering
    \includegraphics{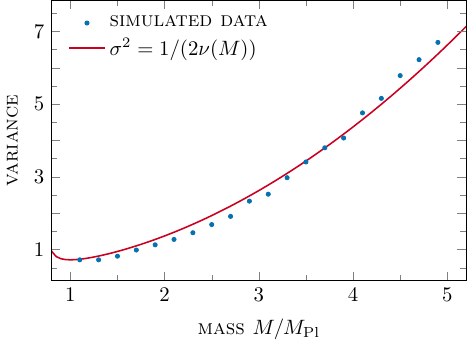}
    \caption{Estimates of \(\nu(M)\) based on spin dynamics simulations (points) versus the analytical expression in Eq.~\eqref{eqn:cm} (blue curve).}
    \label{num}
\end{figure}

To find an expression for \(\nu(M)\), we note that the stationary distribution should satisfy local detailed balance: in equilibrium the probability of observing a transition from \(J\) to \(J + 1\) should be the same as the probability of observing a transition from \(J+1\) to \(J\).  Namely,
\begin{equation}\label{ldb}
    p_\lambda(J) \rho_{\uparrow}(J,M) = p_\lambda(J+1) \rho_{\downarrow}(J+1,M).
\end{equation}
where \(M = \Mbar(\lambda)\).  This in turn implies that
\begin{align}\label{ldb2}
    \frac{p_\lambda(J+1)}{p_\lambda(J)} = \frac{(M^2-J)^2}{M^4+2(J+1)M^2-(J+1)^2}
\end{align}
for \(0 \le J < \lfloor M^2 \rfloor\).
We can thus solve for the successive differences
\begin{equation}\label{eqn:diff}
    E_\lambda(J+1)-E_\lambda(J) = \ln\!\left[ \frac{M^4+2(J+1)M^2-(J+1)^2}{(M^2-J)^2}\right]
\end{equation}
for \(0 \le J < \lfloor M^2 \rfloor\).
To estimate \(\nu(M)\) we only need to know \(E_\lambda(1) - E_\lambda(0) = \nu(M)\), and hence plug in \(J = 0\) into Eq.~\eqref{eqn:diff} to find
\begin{equation}\label{eqn:cm}
    \nu(M) = \ln\!\left[ \frac{M^4+2M^2-1}{M^4} \right].
\end{equation}
Since \(E_{\lambda} (J)\) is not exactly parabolic, different choices of \(J\) would give somewhat different estimates of \(\nu (M)\), but the \(J = 0\) expression is the simplest, and works well compared to the numerics. 
In Fig.~\ref{num} we show \(\nu(M)\) from Eq.~\eqref{eqn:cm} versus numerical estimates (based on the same type of simulations as in Fig.~\ref{fig:parab}).

With proper normalization, the expression for \(p_\lambda(J)\) is given by
\begin{equation}\label{eqn:pt}
    p_\lambda(J) = \frac{e^{-\nu(M)J^2}}{\sum_{J^\prime=-\lfloor M^2 \rfloor}^{\lfloor M^2 \rfloor} e^{-\nu(M){J^\prime}^2}} \approx \sqrt{\frac{\nu(M)}{\pi}} e^{-\nu(M)J^2},
\end{equation}
where we have approximated the sum in the denominator by an integral over \(J^\prime\), letting the bounds go to \(\pm \infty\) since the contribution from the Gaussian tails is negligible.
\newline
\subsection*{Analytical expression for final mass distribution}
\noindent
Putting together Eqs.~\eqref{eqn:pmf}, \eqref{eqn:kt}, and \eqref{eqn:pt}, we get an expression for the final mass distribution
\begin{equation}\label{eqn:final}
\begin{split}
    &P(\Mext)=\\
    &\quad \frac{8\sqrt{\pi}}{\gamma} \Mext \sqrt{\nu (\Mtilde)}
    \left(1 -
        \frac{\lfloor \Mext^2 \rfloor}{\Mtilde^2}
    \right)^2
     e^{-\nu(\stackrel{\sim}{\smash{M}\rule{0pt}{0.9ex}}) \lfloor \Mext^2\rfloor^2},
\end{split}
\end{equation}
where \(\nu(\Mtilde)\) is given by Eq.~\eqref{eqn:cm} and \(\Mtilde \!(\Mext)\) is given by Eq.~\eqref{eqn:Mtilde}. Note that since \(P(\Mext) \,\dd\Mext\) is the probability of a black hole having a final mass between \(\Mext\) and \(\Mext + \dd\Mext\), the integral  \(\int_{1}^{\infty} P(\Mext) \, \dd \Mext \) gives the overall probability of a black hole becoming extremal. The integral can be evaluated numerically to give a value of 21.1\% for this probability.

\end{document}